\documentclass[twocolumn,preprintnumbers,superscriptaddress,nofootinbib]{
revtex4}
\usepackage{graphicx}
\usepackage{amssymb}
\usepackage{color}
\usepackage{enumerate}

\newcommand{\be}{\begin{equation}}
\newcommand{\ee}{\end{equation}}
\newcommand{\bea}{\begin{eqnarray}}
\newcommand{\eea}{\end{eqnarray}}

\begin{document}

\title{One-loop effective potential in nonlocal scalar field models}
\author{F. Briscese}\email{briscese.phys@gmail.com}
\affiliation{ Departamento de F\'isica, Universidade Federal da
Para\'iba, Caixa Postal 5008, 58051-970 Jo\~ao Pessoa, PB, Brazil.
}

\affiliation{Instituto de Investigaciones en Materiales, Universidad Nacional Aut%
\'{o}noma de M\'{e}xico, A. P. 70-360, 04510 M\'{e}xico, DF,
M\'{e}xico. \footnote{FB is visiting researcher at this
institution from October 2015.}}

\affiliation{Istituto Nazionale di Alta Matematica Francesco
Severi, Gruppo Nazionale di Fisica Matematica, Citt\`a
Universitaria, P. le A. Moro 5, 00185 Rome, EU.}

\author{E. R. Bezerra de Mello}\email{emello@fisica.ufpb.br}
\affiliation{ Departamento de F\'isica, Universidade Federal
da Para\'iba, Caixa Postal 5008, 58051-970 Jo\~ao Pessoa, PB,
Brazil. }

\author{A. Yu. Petrov}\email{petrov@fisica.ufpb.br}
\affiliation{ Departamento de F\'isica, Universidade Federal
da Para\'iba, Caixa Postal 5008, 58051-970 Jo\~ao Pessoa, PB,
Brazil. }

\author{V. B. Bezerra}\email{valdir@fisica.ufpb.br}
\affiliation{ Departamento de F\'isica, Universidade Federal
da Para\'iba, Caixa Postal 5008, 58051-970 Jo\~ao Pessoa, PB,
Brazil. }

\begin{abstract}
In this paper we apply the usual perturbative methodology to
evaluate the one-loop effective potential in a nonlocal scalar
field theory. We find that the effect induced by the nonlocaliity
of the theory is always very small and we discuss the consequences
of this result. In particular we argue that, looking at one-loop
corrections for matter fields, it is not possible to find signals
of the nonlocality of the theory in cosmological observables
since, even during inflation when energies are very high,
nonlocality-induced corrections are expected to be very small.
\end{abstract}

\maketitle

\section{Introduction}\label{introduction}

\noindent

Higher-derivative field theory models actually attract great
attention due to two main reasons. The first one is of
cosmological nature. For instance,  cosmological models have been
considered, which are based on the inclusion of finite higher
derivatives (see \cite{Clifton:2011jh} and references therein),
with Lagrangian density $\mathcal{L} = F(R, \Box R, \Box^2
R,\ldots,\Box^m R, \Box^{-} R,\ldots,\Box^{-m} R)$. Such models
are capable to explain inflation and dark energy in a unified
framework \cite{odintsov1,odintsov3} (see also \cite{odintsov2}
for a complete review). The second reason is connected with the
possibility to construct an higher-derivative theory which can be
a viable quantum gravity candidate, together with other approaches
to this scenario, namely, Loop Quantum Gravity, Strings and
Noncommutative Geometries (see \cite{smolin} for a review).

Among the theories addressing questions related to quatum gravity, one has received
a great deal of attention, the so-called Horava-Lifshitz Gravity
(H-LG) \cite{horava}. This assumes a foliation of the spacetime
and is characterized by a Lagrangian density which contains
higher than first order spatial derivatives and is invariant under
foliation-preserving diffeomorphisms. Such a theory is
power-counting renormalizable due to the higher spatial
derivatives which make the graviton propagator to converge to zero
more rapidly than $1/k^2$ at high wave-numbers $k$ \cite{yeinzon}.
Moreover H-LG has been shown to be a viable model of gravitation
at cosmological and astrophysical level \cite{horava}.  Among
different implications of the H-LG, one should mention the consistency with the
well-known gravitational solutions, such as black holes \cite{Kon}, FRW
solutions \cite{FRW}, and G\"{o}del-type metrics \cite{Reb}, and some new cosmological
concepts such as bouncing Universe \cite{Brand} and anisotropic scaling
\cite{anis}, as well as the quantum studies that allow to discuss the possible contruction
of a renormalizable gravity theory(see ref.\cite{quant}).

Despite its advantages, H-LG  involves an
apparently strong problem: the loss of diffeomorphism invariance
and then the loss of Lorentz symmetry in flat spacetime, as
discussed in \cite{pospelov}. To avoid this  problem exploiting
the better convergence of the graviton propagator in higher
derivative theories, one can consider covariant higher derivative
theories of gravitation as in \cite{stelle}, which are invariant
under spacetime diffeomorphisms. The main problem here is that such theories
usually contain a physical
ghost (a state of negative norm) and therefore they violate the
unitarity.  It is worth to mention the papers \cite{CMR}, where
the higher-derivative extensions of different field theory models
were introduced without breaking the unitarity; however, in these
theories the higher derivatives are present only in a term proportional
to Lorentz-breaking parameters which leads to specific effects like
large quantum corrections and fine-tunning.

In order to overcome this problem, a new class of higher derivative
theories which manifestly preserve Lorentz invariance, the Non
Local Quantum Gravity (NLQG) models, has been proposed recently
\cite{modesto1,biswas}.  These models were, partially inspired by \cite{stelle},
and their construction were done in order to fulfill the following hypotheses:
(i) Classical solutions must be singularity-free; (ii)
Einstein-Hilbert action must be a good approximation of the theory
below the Planck energy scale; (iii) The theory must be
perturbatively quantum-renormalizable on a flat background; (iv)
The theory must to be unitary; (v) Lorentz invariance must be
preserved.

Models in the framework of NLQG are both renormalizable and ghost-free
\cite{modesto1,biswas}, and therefore, they have no the shortcomings of
Einstein's gravity related to these points.  The typical Lagrangian density for NLQG
is a non-polynomial extension of the renormalizable quadratic
Stelle theory\cite{stelle} and it has the following structure:

\be \! \mathcal{L} =  R - \left( R_{\mu \nu}- \frac{1}{2} R
g_{\mu\nu} \right) \gamma(\Box/\Lambda^2) R^{\mu \nu} \, ,
\label{theory} \ee where the form factor $\gamma(z)$ is a function
without poles,  $\Box$ is the covariant D'Alembertian operator and
$\Lambda$ is an invariant mass scale, which we expect to be close
to the Planck  mass, since we want that our theory reduces to the
Einstein's theory in the low energy limit. We stress that
$1/\Lambda$ represents the length scale above which the theory is
fully nonlocal. Note that the local behavior of the theory is
recovered at energies below $\Lambda$ and that all degrees of
freedom present in the action can be embedded in the function
$\gamma(\Box/\Lambda^2)$.

It is convenient to express the form factor
$\gamma(\Box/\Lambda^2)$ introducing a new form factor
$V(\Box/\Lambda^2)$ defined as

\be
\gamma(\Box/\Lambda^2) \equiv \frac{V(\Box/\Lambda^2)^{-1}
-1}{\Box}  \label{FFS}  \   ,
\ee
so that the propagator of the theory is

\begin{equation}
G(k^2) = \frac{V(k^2/\Lambda^2)  } {k^2} \left( P^{(2)} -
\frac{P^{(0)}}{2 }  \right)\,, \label{propgauge2}
\end{equation}
where $P^{(0)}$ and $P^{(2)}$ are the spin zero and spin two
projectors \footnote{The projectors are defined by \cite{HigherDG,
VanNieuwenhuizen} $ P^{(2)}_{\mu \nu, \rho \sigma}(k) =
\frac{1}{2} ( \theta_{\mu \rho} \theta_{\nu \sigma} + \theta_{\mu
\sigma} \theta_{\nu \rho} ) - \frac{1}{D-1} \theta_{\mu \nu}
\theta_{\rho \sigma} \, $, $ P^{(1)}_{\mu \nu, \rho \sigma}(k) =
\frac{1}{2} \left( \theta_{\mu \rho} \omega_{\nu \sigma} +
\theta_{\mu \sigma} \omega_{\nu \rho} + \theta_{\nu \rho}
\omega_{\mu \sigma}  + \theta_{\nu \sigma} \omega_{\mu \rho}
\right) \, $, $ P^{(0)} _{\mu\nu, \rho\sigma} (k) = \frac{1}{D-1}
\theta_{\mu \nu} \theta_{\rho \sigma}  \, , \,\,\,\, \,\,
\bar{P}^{(0)} _{\mu\nu, \rho\sigma} (k) =  \omega_{\mu \nu}
\omega_{\rho \sigma} \, $, $ \theta_{\mu \nu} = \eta_{\mu \nu} -
\frac{k_{\mu} k_{\nu}}{k^2} \, , \,\,\,\,\, \omega_{\mu \nu} =
\frac{k_{\mu} k_{\nu}}{k^2} \, $, where $D$ is the dimension of the spacetime}.


Since one wants to recover Einstein's gravity for small momenta,
one needs to impose that $V(0)= 1$ and $V(z) \simeq 1$ for $|z|
\ll 1$. It is now evident that, if $V(z)$ has no poles, the only propagating
degrees of freedom corresponds to the two polarizations of Einstein's theory,
and thus the theory does not contain ghosts. As a consequence, the
function $\gamma(z)$ cannot be polynomial, otherwise the
function $V(z)$ would have a pole at $z_i$. Therefore, in order to avoid ghosts,
we have to pay a price, namely, NLQG must contain
derivatives of arbitrary order, which means that it must be non-local.
Moreover, from (\ref{propgauge2}) it follows that, if $V(z)$ goes to
zero for $|z| \gg 1$ sufficiently fast, the theory is
super-renormalizable, since this requirement improves the
convergence of the integrations over loops (see for instance
\cite{modesto1} for details).

The class of nonlocal theories (\ref{theory}) have nice
properties. It has been shown in \cite{briscese1} that NLQG has
two degrees of freedom corresponding to the spin 2 graviton and
at most one extra scalar degree of freedom: any further extra
degree of freedom in fact would correspond to a ghost or a
tachyon, breaking the unitarity of the theory. Moreove, in
\cite{briscese1,briscese2} it has been shown that, at cosmological
level, NLQG reduces to the $R + \epsilon R^2$ Starobinsky model
\cite{Staro} with the identification $\epsilon \equiv 1/\Lambda^2$
up to corrections of order $1/\Lambda^4$. Therefore, NLQG gives a
viable inflation in agreement with Planck data \cite{planck} for
$\Lambda \sim 10^{-5} \, M_P$ \cite{briscese1,briscese2}.

A further relevant feature of NLQG is that, due to higher
derivatives, it can be free from the singularities which affect
Einstein's gravity. In fact, in NLQG the linearized equations for
gravitational perturbations of Minkowski background typically
reads $\exp(\Box/\Lambda^2) \Box h_{\mu \nu} = M_P^{-2}  \tau_{\mu
\nu}$ \cite{biswas}, where $\tau_{\mu \nu}$ is the stress energy
tensor of matter. For a point-like source of mass m this gives a
Newtonian potential $h_{00} \sim erf(r\Lambda/2)\,  m/r M_P^2$,
where $erf(z)$ is the error function of argument $z$, which is
finite for all $r \geq 0$. This shows how black holes
singularities of general relativity can be removed in NLQG (see
also \cite{modesto3}). With similar arguments one can show that
also the big bang singularity can be removed and a non singular
bouncing cosmology can be obtained \cite{modesto2}.

The problem with NLQG is that, due to the high  complexity of the
theory, it is very difficult to perform explicit calculations.
This is why, in order to understand the basic properties of
nonlocal models, it is interesting to consider toy models, e.g.
nonlocal scalar field theory.  The properties of nonlocal scalar
fields has been first studied by Efimov in a series of seminal
papers \cite{efimovquantization,efimovunitarity,efimovcausality}.
Specifically, the quantization scheme has been described in
\cite{efimovquantization}, the unitarity of the theory has been
demonstrated in \cite{efimovunitarity} and the causality has been
discussed in \cite{efimovcausality}. Here we also mention that a
nonlocal version of QED has been studied in \cite{efimovqed} and
nonlocal vector field theory has been introduced in \cite{tomb}.

More recently, in \cite{biswasscalar} it has been considered the
case of a nonlocal scalar field with specific self--interactions
which have been chosen to present the same symmetries of the NLQG; a
toy model depiction of a non-local gravity. The form factor in
(\ref{FFS}) has been chosen in such a way that $V(z) = \exp(-z)$
and the graviton propagator in (\ref{propgauge2}) has an
exponential suppression, so that it is asymptotically free. In particular,
the 2-point function is still divergent, but it can be renormalized by adding
appropriate counterterms, so that the ultraviolet behavior of all other 1-loop
diagrams as well as the 2-loop, 2-point function remains finite.

In this paper  we  consider a nonlocal scalar field model with
generic potential and calculate the one-loop corrections to this
quantity. We show that the corrections induced by the nonlocality
of the theory, with respect to the local one, are very small even
in the strong field limit, i.e., when the typical energies are of
the order or much greater than $\Lambda$. This implies that it
turns out to be very hard to find traces of the nonlocal nature of
the theory looking at the one-loop corrections to the bare
potential. For instance, one might hope to find non-locality
signatures in very energetic contexts,  for instance in inflation.
If the inflaton field, which is responsible of the
nearly-exponential of the early universe, is actually nonlocal,
one can seek traces of the nonlocality of the theory in
cosmological observables. However, since the one-loop corrections
to the inflaton potential would be very small, it is impossible to
detect the effect of one-loop corrections to the cosmological
observables, even using the most precise measurements currently
available, as the CMB temperature and polarization measurements
made by Planck \cite{planck}. This suggests us that one should
look for the signals of nonlocality studying other physical
effects.

This paper is organized as follows: In section \ref{nonlocal
scalar field} we introduce the nonlocal scalar field action and
discuss its properties. In section \ref{one loop corrections} we
calculate the one-loop correction to the potential for different
nonlocal field actions. Finally we leave to the section \ref{conclusions}
our conclusions and most relevant remarks.

\section{Nonlocal scalar field}
\label{nonlocal scalar field}

In this paper we consider nonlocal scalar fields with a Lagrangian density
given by

\begin{equation}\label{nonlocal action}
{\cal L}=- \frac{1}{2} \phi F(\Box/\Lambda^2) \phi - V(\phi) \  ,
\end{equation}
where $V(\phi)$ is the scalar field potential which describes
self-interactions (for instance $V(\phi)= \lambda \phi^4/4!$) and
$F(z)$ is a non-polynomial analytic function, which in most cases
can be represented by a series expansion

\begin{equation}\label{F(z)}
F(z) \equiv \sum_{n=0}^\infty f_n z^n \   ,
\end{equation}
with $f_n \neq 0$ for any $n> n_0$, with $n_0 \in N$. By
definition $F(\Box/\Lambda^2)$ contains derivatives of arbitrary
order, which makes the theory nonlocal.

From Eq. (\ref{nonlocal action}) it is immediate to recognize that the
free field propagator of the theory is

\begin{equation}
G(k^2) = - \frac{1}{F(-k^2)}
\end{equation}
and this expression makes evident that to each zero of the function
$F(z)$ corresponds a pole in the propagator, and therefore a physical particle.

This conclusion is made more evident by the following
considerations (see \cite{fieldred}). Suppose that $F(z)$ has a
finite number $M$ of zeroes. Due to the Weierstrass factorization
theorem, we can decompose it as

\begin{equation}\label{F(z) factorized}
F(z) = f(z) \prod_{a=1}^M (z \Lambda^2 + m^2_a)^{r_a} \, ,
\end{equation}
where the function $f(z)$  has no zeroes and no poles  and $r_a$
corresponds to a positive interger number. We can therefore use a
field redefinition
\begin{equation}\label{field redefinition}
\varphi \equiv f(\Box/\Lambda^2)^{1/2} \phi \  ,
\end{equation}
so that the Lagrangian given by Eq. (\ref{nonlocal action}) becomes

\begin{equation}\label{nonlocal action 2}
{\cal L}=- \frac{1}{2}\varphi \prod_{a=1}^M (\Box + m^2_a)^{r_a}
\varphi - V\left(\Gamma(\Box/\Lambda^2)^{-1/2} \, \varphi\right).
\end{equation}

At this point one can introduce the M independent fields
$\varphi$ as

\begin{equation}\label{varphi a}
\varphi^a \equiv \prod_{b\neq a} (\Box/\Lambda^2 + m^2_b)^{r_b}
\varphi, \quad\, a=1,2\ldots M .
\end{equation}
and express the field $\phi$ as a superposition of the fields
$\varphi^a$ as

\begin{equation}\label{phi 1}
\phi = f(\Box/\Lambda^2)^{-1/2} \sum_{a=1}^M \eta_a \varphi^2
\, ,
\end{equation}
where the coefficients $\eta_a$ satisfy the relation
\begin{equation}\label{eta definition}
\sum_{a=1}^M \frac{\eta_a}{(z +m_a^2)^{r_a}} = \left(\prod_{a=1}^M
(z + m_a^2 )^{r_a} \right)^{-1} \  .
\end{equation}

Therefore the Lagrangian density  (\ref{nonlocal action 2}) can be
expressed as

\begin{equation}\label{nonlocal action 3}
{\cal L}=- \sum_{a=1}^M \frac{1}{2} \eta_a \varphi^a (\Box +
m^2_a)^{r_a} \varphi^a - V\left(\Gamma(\Box/\Lambda^2)^{-1/2}
\sum_{b=1}^M \eta_b \varphi^b \right) .
\end{equation}

Restricting to the case in which all zeroes of $F(z)$ are
simple, i.e., $r_a = 1 \,\,\, \forall a$, one has that the theory
contains M interacting constituent scalar fields. Since the
$\eta_a$ have alternating sign, thus some of these
constituent fields are ghosts. We also note the following: from
Eq. (\ref{nonlocal action 3}) it is evident that, in the new
variables, the nonlocality of the theory is contained only in the
potential $V$, and therefore it follows that a non-interacting
nonlocal theory is in fact local.

Since from the previous considerations it follows that the unique
ghost free case is that with $M=1$, from now on we limit our interest
to functions of the type $F(z) = f(z) \left(z-m^2 \right)$,
so that

\begin{equation}\label{nonlocal action 4}
{\cal L}=- \frac{1}{2}\phi f(\Box/\Lambda^2) (\Box + m^2)
\phi - V(\phi)   \  ,
\end{equation}
from which we can write the propagator, which is given by

\begin{equation}\label{propagator}
G(k^2) = \frac{1}{f(-k^2/\Lambda^2) \left( k^2 - m^2 \right)}
.
\end{equation}

From the knowledge of the propagator it is possible to obtain the
one-loop correction to the scalar field potential, as is explained
in the next section

\section{One-loop corrections}
\label{one loop corrections}

In this section we calculate the one-loop effective potential for this theory.
To do it, we generalize the formula (9-119) of \cite{Itzkynson},
to find  the one-loop correction to the scalar field potential

\begin{equation}\label{oneloop 1}
\begin{array}{ll}
V^{(1)}= -\frac{i}{2} \int\frac{d^4k}{(2\pi)^4}\left[\ln\left[1-
G(k^2) V^{\prime\prime}\right] +  \, G(k^2) V^{\prime\prime}
+\right.\\
\\
\left. +\frac{1}{2} \left( G(k^2) V^{\prime\prime} \right)^2
\right] \  .
\end{array}
\end{equation}

The second and third terms in (\ref{oneloop 1}) are added, as in
the standard local case, to comply with the prescription of normal
ordering, which avoid the inclusion of tadpole diagrams, and
corresponds to the-one-loop diagrams of the two- and four-point
functions \cite{Itzkynson}.

Performing a Wick rotation, we obtain
\begin{equation}\label{oneloop 2}
\begin{array}{ll}
V^{(1)}= \frac{1}{2} \int\frac{d^4k_E}{(2\pi)^4}\left[\ln\left[1-
G(-k_E^2) V^{\prime\prime}\right] + G(-k_E^2) V^{\prime\prime}
+\right.\\
\\
\left. + \frac{1}{2} \left( G(-k_E^2) V^{\prime\prime} \right)^2
\right] \   .
\end{array}
\end{equation}

Therefore, to calculate the one-loop correction to the potential
we only need to know the form of the propagator $G(k^2)$. In what
follows we make this calculation explicitly in the case of some
specific choice of the function $F(z)$ which reduces to the local model at $\Lambda\to\infty$.

\subsection{Case 1}

We firstly consider the theory described by the Lagrangian density

\begin{equation} {\cal L}=- \frac{1}{2}\phi
\left(\exp(\Box/\Lambda^2)\Box + m^2\right)\phi - V(\phi) \, ,
\end{equation}
where we have considered the same exponential nonlocal factor used
in \cite{biswasscalar}. This Lagrangian provides the scalar
propagator

\begin{equation}\label{propagator 1}
G(k^2) =  \frac{1}{ k^2 \exp\left(-k^2/\Lambda^2\right) - m^2}   \   .
\end{equation}

From Eq.(\ref{oneloop 2}) and after integration over the solid
angle, one has

\begin{equation}\label{v1 0}
V^{(1)}= \frac{1}{(4\pi)^2} \int_0^\infty dk_E \,
F(k_E,\Lambda,m,V'')\   ,
\end{equation}
where
\begin{equation}
\begin{array}{ll}
F(k_E,\Lambda,m,V'') \equiv k_E^{3} \left[ \ln\left[1+
\frac{V^{\prime\prime}}{k_E^2 \exp[k_E^2/\Lambda^2]+m^2}\right]+\right.\\
\\
\left.  - \frac{V^{\prime\prime}}{k_E^2 \exp[k_E^2/\Lambda^2]+m^2}
\frac{1}{2} \left( \frac{V^{\prime\prime}}{k_E^2
\exp[k_E^2/\Lambda^2]+m^2} \right)^2 \right] \   ,
\end{array}
\end{equation}

From the last expression we deduce the following: the departure
from the standard local result is given by integration at high
momenta $k_E \gtrsim \Lambda$, where the exponential
$\exp(k_E^2/\Lambda^2)$ is significatively different from unity.
Therefore, if $V^{\prime\prime}/(k_E^2 \exp(k_E^2/\Lambda^2)+m^2)\ll
1$ at such high momenta, the integrand is very small and the
effect of nonlocality is negligible. Thus we expect two
significantly different regimes: the strong field regime $V''
\gtrsim \Lambda$, where the effect of nonlocalities is stronger,
and the weak field regime $V''\ll \Lambda$, where a similar effect
is weaker.

Before proceeding the explicit calculation of (\ref{v1 0}), let
us recall the one-loop correction in the case of a local model.
The local model is obtained in the limit $\Lambda \rightarrow
\infty$, and it is given by

\begin{equation}
\label{v1 local}
\begin{array}{ll}
V^{(1) }_{local} = \frac{1}{(4\pi)^2} \int_0^\infty  dk_E \,
F_\infty(k_E,m,V'') = \\
\\
=\frac{1}{(8\pi)^2} \left[ \left(V'' + m^2 \right)^2 \ln\left[ 1 +
\frac{V''}{m^2}\right] -  V'' \left( \frac{3}{2} V'' + m^2 \right)
\right],
\end{array}
\end{equation}
where we have defined
\begin{equation}
\label{F infty} \begin{array}{ll} F_\infty(k_E,m,V'') =
\lim\limits_{\Lambda \rightarrow \infty} F(k_E,\Lambda,m,V'')  =\\
\\
k_E^{3} \left[ \ln\left[1+ \frac{V^{\prime\prime}}{k_E^2
+m^2}\right] - \frac{V^{\prime\prime}}{k_E^2 +m^2} + \frac{1}{2}
\left( \frac{V^{\prime\prime}}{k_E^2 +m^2} \right)^2 \right]   \ .
\end{array}
\end{equation}

It is useful for our proposal, to express Eq. (\ref{v1 0}) as
\begin{equation}\label{v1 1}
\begin{array}{ll}
V^{(1)} \simeq  \frac{1}{(4\pi)^2} \int_\Lambda^\infty dk_E \,
\left[ F_0(k_E,\Lambda,V'') -
F_\infty(k_E,m,V'') \right] + \\
\\+ V^{(1)}_{local} = V^{(1)}_{local} + \delta V^{(1)}
\end{array}
\end{equation}
where
\begin{equation}\label{v1 2}
\delta V^{(1)} \equiv  V^{(1)}_{a} + V^{(1)}_{b}
\end{equation}
represents the deviation from the result obtained in the local
theory and where we have defined

\begin{equation}
\begin{array}{ll}
\label{v1 a} V^{(1)}_a \equiv -\frac{1}{(4\pi)^2}
\int_\Lambda^\infty dk_E \, F_\infty(k_E,m,V'') =\\
\\
= - \frac{1}{(8 \pi)^2} \left[ \ln\left[1+
\frac{V''}{\Lambda^2+m^2}\right] \times  \left( V''\left(V''+ 2
m^2 \right)+ m^4-\Lambda^4\right) \right. \\
\\
\left.-\frac{V''}{2\left(\Lambda^2+m^2 \right) }\left(3 V'' m^2 +
2 m^4 + V''\Lambda^2-2\Lambda^4\right) \right]
\end{array}
\end{equation}
and

\begin{equation}\label{v1 b}
V^{(1)}_b \equiv \frac{1}{(4\pi)^2} \int_\Lambda^\infty dk_E \,
F_0(k_E,\Lambda,V'') \  ,
\end{equation}
with
\begin{equation}
\begin{array}{ll}
F_0(k_E,\Lambda,V'') \equiv F(k_E,\Lambda,m=0,V'') =\\
\\
= k_E^{3} \left[ \ln\left[1+ \frac{V^{\prime\prime}}{k_E^2
\exp[k_E^2/\Lambda^2] }\right]- \frac{V^{\prime\prime}}{k_E^2
\exp[k_E^2/\Lambda^2] } + \right.\\
\\
+ \left. \frac{1}{2} \left( \frac{V^{\prime\prime}}{k_E^2
\exp[k_E^2/\Lambda^2] } \right)^2 \right].
\end{array}
\end{equation}
While $V^{(1)}_a $ is calculated exactly, it is not possible to
find an analytic expression for $V^{(1)}_b$, and we have to
perform its expression by making some simple assumption, namely, considering the
strong and weak field limits.

\subsubsection{Strong field limit}

Let us proceed with the calculation of (\ref{v1 b}) in the strong
field limit $V''\gg \Lambda^2$. First of all let us divide the
integration region in two intervals: $I_1\equiv [\Lambda, \bar{k}]$ and
$I_2 \equiv [\bar{k}, \infty [$, where $\bar{k} \equiv  \Lambda
\sqrt{|X|} $ and X is the solution of the equation $X \exp[X] =
V''/\Lambda^2$, which is approximately given by

\begin{equation}\label{X 0}
X \simeq \ln \left[ V''/\Lambda^2 \right] - \ln \left[ \ln
\left[V''/\Lambda^2\right] \right].
\end{equation}
Therefore, since $V''/\Lambda \gg 1$ one has $\bar{k} > \Lambda$.

The contribution to $V^{(1)}$ coming from the integration over
$I_1$ is therefore
\begin{equation}\label{v1 I2 2}
\begin{array}{ll}
V^{(1)}_{I_1}(\bar{k})\simeq \frac{1}{(4\pi)^2}
\int_\Lambda^{\bar{k}} dk_E \, F_0(k_E,\Lambda,V'') = \\
\\
= \frac{1}{6 (8\pi)^2} \left[ 6 V''^2 \left(EI\left[- 2
\bar{k}^2/\Lambda^2\right]- EI\left[-2\right] \right)
+\right.\\
\\
\left. +3\left(\bar{k}^4 - \Lambda^4 \right) + 2 \left(\bar{k}^6 -
\Lambda^6\right)/\Lambda^2 + \right.\\
\\
+ \left. 12 V'' \Lambda^2
\left(\exp\left[-\bar{k}^2/\Lambda^2\right] - 1 \right)  - 6
\Lambda^4 \ln\left[V''/\Lambda^2 \right] \right]
\end{array}
\end{equation}
where  $Ei(z) \equiv -\int_{-z}^\infty dt \exp[-t]/t$ is the
exponential integral function and $Ei(-2)\simeq -0.0489$. Note
that in the calculation we have neglected $1$ with respect to
$V''/k_E^2 \exp[k_E^2/\Lambda^2]$.

Let us calculate the contribution arising from the integration over
$I_2$. Since in this region $V''/k_E^2 \exp[k_E^2/\Lambda^2] \ll
1$, we can expand the logarithm in $F_0(k_E,\Lambda,V'')$ in power
series and then integrate, so that we obtain

\begin{equation}\label{v1 I3}
\begin{array}{ll}
V^{(1)}_{I_2}(\bar{k})\simeq \frac{1}{(4\pi)^2}
\int_{\bar{k}}^\infty dk_E \, F_0(k_E,\Lambda,V'') = \\
\\
=\Lambda^4 \sum_{n = 3}^\infty C_n(\bar{k})
\left(\frac{V''}{\Lambda^2}\right)^n  \  ,
\end{array}
\end{equation} where

\begin{equation}\label{Cn}
C_n(\bar{k}) \equiv \frac{(-1)^{n+1}}{2(4 \pi)^2} \frac{1}{
n^{3-n}} \Gamma(2-n, n \bar{k}/\Lambda^2)
\end{equation}
with $\Gamma(a,x)$ being the incomplete Gamma function. Therefore one has

\begin{equation}\label{v1 b1}
V^{(1)}_b  = V^{(1)}_{I_1}(\bar{k}) + V^{(1)}_{I_2}(\bar{k}) \   .
\end{equation}

To evaluate the effect of non-locality we can estimate the value
of $\delta V^{(1)}/V^{(1)}_{local}$ in the strong field limit.
From (\ref{v1 local}), (\ref{v1 I2 2}), (\ref{v1 I3}) and  (\ref{v1 a}),  one has

\begin{equation}\label{estimation Vlocal}
V^{(1)}_{local} \simeq \left(\frac{V''}{8\pi}\right)^2 \ln\left[1+
V''/m^2 \right],
\end{equation}

\begin{equation}\label{estimation Va}
V^{(1)}_{a} \simeq - \left(\frac{V''}{8\pi}\right)^2 \ln\left[1+
V''/\Lambda^2 \right],
\end{equation}

\begin{equation}\label{estimation I2}
V^{(1)}_{I_2} \simeq \left(\frac{V''}{8\pi}\right)^2
|Ei\left[-2\right]| \   .
\end{equation}
So,
\begin{equation}\label{estimation I3}
V^{(1)}_{I_3} < \left(\frac{V''}{8\pi}\right)^2 \ln\left[1+
V''/\bar{k}^2 \right] \simeq \left(\frac{V''}{8\pi}\right)^2
\ln\left[1+ V''/\Lambda^2 \right]
\end{equation}

The estimations given by Eqs. (\ref{estimation Vlocal}), (\ref{estimation
Va}) and  (\ref{estimation I2}) are obtained by taking the limit $V'' \gg
\bar{k}^2 \gg \Lambda^2\gg m^2$ in equations (\ref{v1
local}), (\ref{v1 I2 2}), (\ref{v1 I3}) and  (\ref{v1 a}). To obtain
(\ref{estimation I3}) it is necessary to note that, since the
function $g(x) = \ln[1+x]-x +x^2/2$ grows monotonically for any $x
> -1$, one has

\begin{equation}\label{estimation I3 2}
\begin{array}{ll}
V^{(1)}_{I_3} < \frac{1}{(4\pi)^2} \int_{\bar{k}}^\infty dk_E
\,k_E^3 \left[\ln\left[1+ V''/k_E^2
\right]+ \right.\\
\\
- \left. V''/k_E^2+(V''/k_E^2)^2/2 \right]  =
\frac{\bar{k}^4}{(8\pi)^2} \left[ V''/\bar{k}^2
\left(1-V''/2\bar{k}^2 \right)   + \right.\\
\\
+ \left. \ln\left[1 +
V''/\bar{k}^2\right]\left((V'')^2/\bar{k}^4-1 \right)  \right] \ .
\end{array}
\end{equation}

Therefore, from Eqs. (\ref{estimation Vlocal}), (\ref{estimation
Va}), (\ref{estimation I2}) and (\ref{estimation I3}) it follows that
$\delta V^{(1)} \simeq V^{(1)}_a$ and therefore

\begin{equation}\label{deltaV V strong}
\frac{\delta V^{(1)}}{V^{(1)}_{local}} \simeq \frac{\ln\left[1+
V''/\Lambda^2 \right]}{\ln\left[1+ V''/m^2 \right]} \ll 1,
\end{equation}
\noindent
which gives a measure of the effect of the non-locality of the
scalar field on the one-loop corrections of the bare potential.
The conclusion is that, even in the strong field limit, since
$\Lambda \gg m$, one has $\delta V^{(1)}/V^{(1)}_{local} \ll 1$, and therefore,
this effect is very small.

\subsubsection{Weak field limit}

Let us consider now the weak field limit in which $V''/\Lambda^2
\ll 1$ but $V'' \gg m^2$. Since in this limit one has $V''/k_E^2
\exp[k_E^2/\Lambda^2] \ll 1$ for any $k_E \geq \Lambda$, one can
always expand the logarithm in (\ref{v1 b}) in power series and
then integrate, so that one obtains

\begin{equation}\label{v1 I3a}
\begin{array}{ll}
V^{(1)}_{b}(\bar{k})\simeq \frac{1}{(4\pi)^2}
\int_{\bar{k}}^\infty dk_E \, F_0(k_E,\Lambda,V'') =\\
\\
= \Lambda^4 \sum_{n = 3}^\infty C_n(\Lambda)
\left(\frac{V''}{\Lambda^2}\right)^n   \   ,
\end{array}
\end{equation}
where

\begin{equation}\label{Cna}
C_n(\Lambda) \equiv \frac{(-1)^{n+1}}{2(4 \pi)^2} \frac{1}{
n^{3-n}} \Gamma(2-n, n)  \   .
\end{equation}

Now, taking the limit $ \Lambda^2 \gg V'' \gg m^2$ in (\ref{v1
local}), ( \ref{v1 a}) and  (\ref{estimation I3 2}),  we obtain the result given
by Eq. (\ref{estimation Vlocal}) and for $V^{(1)}_{a}$, we get

\begin{equation}\label{estimation Va weak}
V^{(1)}_{a} \simeq - \frac{2}{3} \left(\frac{1}{8\pi}\right)^2
\frac{V''^3}{\Lambda^2}.
\end{equation}
So,
\begin{equation}\label{estimation I3 weak}
V^{(1)}_{I_3} < \left(\frac{1}{8\pi}\right)^2
\frac{V''^3}{\Lambda^2} 2 \Gamma(-2,3),
\end{equation}
which finally gives

\begin{equation}\label{deltaV V weak}
\frac{\delta V^{(1)}}{V^{(1)}_{local}}  \simeq
\frac{V''}{\Lambda^2} \frac{1}{\ln[V''/m^2]} \ll 1  \   .
\end{equation}

The comparison of (\ref{deltaV V weak}) with (\ref{deltaV V
strong}) shows that the relative correction to the result
obtained in the local theory is much stronger in the strong field
limit.

\subsection{Case 2}

A second example is easily obtained considering the Lagrangian density

\bea {\cal L}=- \frac{1}{2}\left[\phi
T(\Box)\left(\Box+m^2\right)\phi + V(\phi) \right] \   ,
\eea
where we can choose $T(\Box)= \exp[-\Box/\Lambda^2]$. Calculations
are the same as in the previous case. In fact the one-loop
correction is given by

\begin{equation}\label{v1 0 2}
V^{(1)}= \frac{1}{(4\pi)^2} \int_0^\infty dk_E \,
G(k_E,\Lambda,m,V'') \  ,
\end{equation}
where

\begin{equation}
\begin{array}{rr}  G(k_E,\Lambda,m,V'') \equiv k_E^{3} \left[ \ln[1+ \frac{V^{\prime\prime}\exp[-k_E^2/\Lambda^2]}{k_E^2
+m^2}] + \right.\\
\\
- \left. \frac{V^{\prime\prime}\exp[-k_E^2/\Lambda^2]}{k_E^2 +m^2}
+ \frac{1}{2} \left(
\frac{V^{\prime\prime}\exp[-k_E^2/\Lambda^2]}{k_E^2 +m^2}
\right)^2 \right].
\end{array}\end{equation}

From the simple observation that

\begin{equation}\label{G infty}
\begin{array}{ll}
G_\infty(k_E,m,V'')= \lim_{\Lambda \rightarrow \infty}
G(k_E,\Lambda,m,V'')= \\
\\
=F_\infty(k_E,m,V''),\\
\\
G_0(k_E,\Lambda,V'') \equiv G(k_E,\Lambda,m=0,V'')=
F_0(k_E,\Lambda,V'')
\end{array}
\end{equation}
\noindent
one immediately realizes that one can repeat the same steps of the
previous section and arrive to the same results, especially to
Eqs. (\ref{deltaV V strong}) and (\ref{deltaV V weak}), for strong and weak approximations.

\subsection{Case 3}

A further example in which we expect a different result is given
by the non-local Lagrangian density

\bea {\cal L}=- \frac{1}{2}\left[\phi
T(\Box)\left(\Box+m^2\right)\phi + V(\phi) \right], \eea
where the factor $T(\Box)$ was introduced in
\cite{modesto1}, and is given by

\bea\label{form factor leonardo} T(\Box) \equiv
\exp\left[H(-\Box/\Lambda^2) \right], \eea
where

\begin{equation}
\begin{array}{ll}
H(z) \equiv \frac{1}{2} \left[ \gamma_E + \ln\left[p_n(z)^2\right]
- Ei(-p_n(z)^2) \right] =\\
\\
= \sum_{k=1}^\infty \frac{(-1)^{k+1}}{2 k k!}(p_n(z))^{2k},
\end{array}\end{equation}
and $p_n(z)$
is a polynomial of order $n$ such that $p_n(z) \ll 1$, for any
$z\lesssim 1$ and $p_n(z) \simeq z^n$, for $z\gtrsim 1 $. These
conditions implies that $H(z) \simeq 1$ for $z\lesssim 1$ and
$H(z) \simeq z^{n}$ for $z \gtrsim 1$.

In this case the one-loop correction is given by

\begin{equation}\label{v1 0 1}
V^{(1)}= \frac{1}{(4\pi)^2} \int_0^\infty dk_E
Q(k_E,\Lambda,m,V'')  \   ,
\end{equation}
where now
\begin{equation}
\begin{array}{ll}  Q(k_E,\Lambda,m,V'') \equiv k_E^{3} \left[ \ln\left[1+
\frac{V^{\prime\prime}}
{\exp\left[H\left(k_E^2/\Lambda^2\right)\right]\left(k_E^2
+m^2\right)} \right] + \right.\\
\\
- \left.
\frac{V^{\prime\prime}}{\exp\left[H\left(k_E^2/\Lambda^2\right)\right]\left(k_E^2
+m^2\right)} + \frac{1}{2} \left(
\frac{V^{\prime\prime}}{\exp\left[H\left(k_E^2/\Lambda^2\right)\right]\left(k_E^2
+m^2\right)} \right)^2 \right].
\end{array}\end{equation}

It is worth noting that in this case, we have
\begin{equation}\label{Q infty}
\begin{array}{ll}
Q_\infty(k_E,m,V'')= \lim_{\Lambda \rightarrow \infty}
G(k_E,\Lambda,m,V'')= \\
\\
= F_\infty(k_E,m,V''),\\
\\
G_0(k_E,\Lambda,V'') \equiv G(k_E,\Lambda,m=0,V'') \neq
F_0(k_E,\Lambda,V''),
\end{array}
\end{equation}
and thus, we expect different results compared with previous cases.

The one-loop correction can be expressed again as in Eq. (\ref{v1 1}),
where now
\begin{equation}\label{v1 2a}
\delta V^{(1)} \equiv  V^{(1)}_{a} + V^{(1)}_{b}  \   ,
\end{equation}
and $V^{(1)}_a \equiv -\frac{1}{(4\pi)^2} \int_\Lambda^\infty dk_E
Q_\infty(k_E,m,V'') $ is the same as in (\ref{v1 a}), while

\bea\label{v1 b caso 3}
V^{(1)}_b \equiv \frac{1}{(4\pi)^2}
\int_\Lambda^\infty dk_E Q_0(k_E,\Lambda,V'')  \   ,
\eea
must be calculated.

In what follows we consider the two previous limiting cases.

\subsubsection{Strong field limit}

In the limit $V'' \gg \Lambda^2$, once again, we can divide the integral
(\ref{v1 b caso 3}) into two parts:

\bea\label{v1 I2 3} V^{(1)}_{I_2} \equiv \frac{1}{(4\pi)^2}
\int_\Lambda^{\tilde{k}} dk_E Q_0(k_E,\Lambda,V'')
\eea
and
\bea\label{v1 I3 3} V^{(1)}_{I_3} \equiv \frac{1}{(4\pi)^2}
\int_{\tilde{k}}^\infty dk_E Q_0(k_E,\Lambda,V'') \, ,
\eea
where the $\tilde{k}$ is defined by the relation $V'' \Lambda^{2n}
/\tilde{k}^{2+2n} = 1$. One has

\begin{equation}
\begin{array}{ll}\label{v1 I2 3 result}
V^{(1)}_{I_2} = \frac{1}{2(8\pi)^2} \left[4 \frac{V''
\Lambda^2}{n-1} \left[ \left(
\frac{\Lambda^2}{V''}\right)^{n-1/n+1} -1 \right] + \right.\\
\\
+ \left. \frac{V''^2 }{n} \left[ 1-  \left(
\frac{\Lambda^2}{V''}\right)^{2n/n+1} \right] +  \left(n+1\right)
\left(\tilde{k}^4-\Lambda^4\right) + \right.\\
\\
- \left. \Lambda^4
\ln\left[\left(\frac{V''}{\Lambda^2}\right)^2\right] \, \right]
\simeq \frac{1}{2n} \left(\frac{V''}{8 \pi}\right)^2
\end{array}
\end{equation}
and

\begin{equation}
\begin{array}{ll}\label{v1 I2 3 resulta}
V^{(1)}_{I_3} = C_{\#} \left(\frac{\Lambda^2}{8\pi}\right)^2\left(
\frac{V''}{\Lambda^2} \right)^{2/n+1} \   ,
\end{array}
\end{equation}
where $ C_{\#} = \sum_{m= 3}^\infty
\frac{(-1)^{m+1}}{m\left[\frac{m}{2}\left(n+1\right)-1 \right]} >
0$ is a real number.

As a conclusion, we can say that the estimation given by Eq. (\ref{deltaV V strong}) remains
valid also for the form factor (\ref{v1 I2 3 resulta}).

\subsubsection{Week field limit}

In the week field limit $V''\ll \Lambda^2$, one has

\begin{equation}
\begin{array}{ll}\label{v1 I2 3 weak result}
V^{(1)}_{b} = \frac{\Lambda^4}{(8 \pi)^2}   \sum_{m= 3}^\infty
\frac{(-1)^{m+1}}{m\left[\frac{m}{2}\left(n+1\right)-1 \right]}
\left( \frac{V''}{\Lambda^2}  \right)^m \  ,
\end{array}
\end{equation}
and therefore, we obtain the same resultsas in Eq. (\ref{deltaV V weak}).

\section{Conclusions}\label{conclusions}

In this paper we have considered nonlocal scalar field models with
generic potential and studied the effect of the nonlocality on the
one-loop corrections to the bare scalar field potential.
Three different models have been considered. All of them reduce themselve
to the local model for the case where $\Lambda\to\infty$. Moreover,
in order to estimate the influence of the nonlocality we have considered two different
situations: the strong and weak field limits.
We have found that the nonlocality effect is stronger in the
strong field limit, i.e.,  $V'' \gg
\Lambda^2$, but indeed very small.

The implications of this finding are important and suggest us to
ask the question whether non-locality of fundamental fields might
have observable cosmological signatures.  In fact, one might
suppose that the inflaton field is non-local, so that during
inflation, at very high energies, corrections due the scalar field
potential associated with non-locality might play some role in the
generation of cosmological perturbations. However, due to the
extreme smallness of such corrections, we expect that their trace
in cosmological observables should not be detected.

Let us make this claim more clear and let us consider the slow
roll parameters $\epsilon$ and $\eta$, the power spectrum of
curvature perturbations $P_\zeta(k)$ and the spectral tilt $n(k)$.
These quantities are determined by the complete effective
potential $V_{eff}$ of the inflaton field, and their expressions
are

\begin{equation}
\epsilon = \frac{M_{Pl}^2}{2} \left( \frac{V'_{eff}}{V_{eff}}
\right)^2 \, ,
\end{equation}

\begin{equation}
\eta = M_{Pl}^2 \, \frac{V''_{eff}}{V_{eff}} \, ,
\end{equation}

\begin{equation}
P_\zeta(k) = \frac{1}{24 \pi^2 M_{Pl}^4} \,
\frac{V_{eff}}{\epsilon} \, ,
\end{equation}
and

\begin{equation}
n(k)-1 = - 6 \epsilon + 2 \eta \, ,
\end{equation}
where $M_{Pl}$ is the Planck mass (See \cite{Lyth}).

In the case of a non-local inflaton field, the effective scalar
field potential will be

\begin{equation}
V_{eff} = \frac{m^2}{2} \phi^2 + V(\phi) + V^{(1)}_{local} +
\delta V^{(1)} = V^{local}_{eff} + \delta V^{(1)}
\end{equation}
where $V^{local}_{eff} \gg \delta V^{(1)}$ and again $\delta
V^{(1)} $ represents the non-locality-induced corrections to the
effective inflaton potential. Since in our case one has that
\begin{equation}
\delta V^{(1)} /V^{local}_{eff}
\lesssim \left(V''/\Lambda^2\right)/\ln(1+V''/m^2),
\end{equation}
is extremely small (assuming $\Lambda\sim M_{Pl}$) and since
$\epsilon$, $\eta$,
$P_\zeta(k)$ and $n(k)$ are measured with
nearly-percent precision \cite{planck}, one concludes that there
is no chance to detect the effect of $\delta V^{(1)}$ with current
experimental precision. This suggests that one should look for
other physical effects in order to detect signals of the non-local
nature of the physical fields.

It is interesting to note that the methodology of so-called
coherent states \cite{Spal} proposed as a manner to implement the
noncommutativity alternative to the well known Moyal product
approach, effectively represents itself as an equivalent
description of the nonlocal field theory discussed in this paper.
Therefore, as a by-product of our study, we arrive at some
justification for the coherent states approach.

{\bf Acknowledgements.} This work was partially supported by
Conselho Nacional de Desenvolvimento Cient\'{\i}fico e
Tecnol\'{o}gico (CNPq). FB thanks CNPq for grant 400.237/2014-8
and PAPIIT-UNAM for grant IN100314.


\newpage

\begin{thebibliography}{99}


\bibitem{Clifton:2011jh}
  T.~Clifton, P.~G.~Ferreira, A.~Padilla and C.~Skordis,
  Phys.\ Rept.\  {\bf 513}, 1 (2012), arXiv:1106.2476.

\bibitem{odintsov1}
  S. ~Nojiri and S.~D.~Odintsov,
  Phys.\ Lett.\ B {\bf 659}, 821 (2008), arXiv:0708.0924.

\bibitem{odintsov3}
S.~Capozziello, E.~Elizalde, S.~'i.~Nojiri and S.~D.~Odintsov,
  Phys.\ Lett.\ B {\bf 671}, 193 (2009),
arXiv:0809.1535.

\bibitem{odintsov2}
  S. ~Nojiri and S.~D.~Odintsov,
  Phys.\ Rept.\  {\bf 505}, 59 (2011), arXiv:1011.0544.


\bibitem{smolin} L. Smolin, {\it Three roads to quantum gravity},
London, UK: Weidenfeld \& Nicolson (2000).



\bibitem{horava} P. Ho\v{r}ava, Phys. Rev. D {\bf 79}, 084008 (2009), arXiv: 0901.3775;
 A. E. Gumrukcuoglu, S. Mukohyama, and A. Wang, Phys. Rev. D {\bf 85},
064042 (2012), arXiv: 1109.2609;  P. Ho\v{r}ava and C. M. Melby-Thompson, Phys. Rev.
D {\bf 82}, 064027 (2010), arXiv: 1007.2410; C.~Bogdanos and E.~N.~Saridakis,
Class.\ Quant.\ Grav.\  {\bf 27}, 075005 (2010); arXiv:0907.1636;
  Y.~-F.~Cai and E.~N.~Saridakis,
  JCAP {\bf 0910}, 020 (2009), arXiv:0906.1789.

\bibitem{yeinzon} F. Briscese \textit{et al.}, Found. Phys. \textbf{42}, 1444 (2012), arXiv: 1205.1722.

\bibitem{Kon} R. Konoplya, Phys. Lett. B {\bf 679}, 499 (2009), arXiv: 0905.1523;
R.-C. Cai, L.-M. Cao, N. Ohta, Phys. Rev. D80, 024003 (2009), arXiv: 0904.3670.

\bibitem{FRW} G. Calcagni, JHEP 0909, 112 (2009), arXiv: 0904.0829; E. Kiritsis, G. Kofinas,
Nucl. Phys. B 821, 467 (2009), arXiv: 0904.1334; H. Lu, J. Mei, C. N. Pope, Phys.
Rev. Lett. 103, 091301 (2009), arXiv: 0904.1595.

\bibitem{Reb} J. B. Fonseca-Neto, A. Yu. Petrov, M. J. Reboucas, Phys.
Lett. B {\bf 725}, 412 (2013), arXiv: 1304.4675.

\bibitem{Brand} R. Brandenberger, Phys. Rev. D 80, 043516 (2009), arXiv: 0904.2835.

\bibitem{anis} S. Mukohyama, JCAP 0906, 001 (2009), arXiv: 0904.2190.

\bibitem{quant} G. Giribet, D. L. Nacir, and F. D. Mazzitelli,
JHEP 1009, 009 (2010), 1006.2870; F. D. Mazzitelli, Int. J. Mod. Phys. D {\bf 20},
745 (2011); D. L. Nacir, F. D. Mazzitelli, and L. G. Trombetta,
Phys. Rev. D 85, 024051 (2012), arXiv: 1111.1662.

\bibitem{pospelov} M. Pospelov and Y. Shang, Phys. Rev. D {\bf 85}, 105001 (2012), arXiv: 1010.5249.

\bibitem{stelle} K. S. Stelle, Phys. Rev. D {\bf 16}, 953 (1977).

\bibitem{CMR} C. M. Reyes, Phys. Rev. D87, 125028 (2013), arXiv: 1307.5340; M. Maniatis, C. M. Reyes, Phys. Rev. D89, 056009 (2014), arXiv: 1401.3752.


\bibitem{modesto1} L. Modesto, Astron. Rev. 8.2, 4 (2013); Phys.
Rev. D86 (2012) 044005, arXiv: 1107.2403;  L. Modesto, L. Rachwal,
Nucl.Phys. B {\bf 889} (2014) 228-248, arXiv:1407.8036 [hep-th].


\bibitem{biswas} T. Biswas \textit{et al.},  Phys. Rev. Lett. {\bf 108}, 031101
(2012), arXiv: 1110.5249.

\bibitem{HigherDG}
A.~Accioly, A.~Azeredo and H.~Mukai,
J.\ Math.\ Phys.\  {\bf 43}, 473 (2002).

\bibitem{VanNieuwenhuizen}
P.~Van Nieuwenhuizen,
Nucl.\ Phys.\ B {\bf 60}, 478 (1973).

\bibitem{briscese1} F. Briscese \textit{et al.}
Phys. Rev. D {\bf 89}, 024029 (2014), arXiv: 1308.1413.

\bibitem{briscese2} F. Briscese \textit{et al.}, Phys. Rev. D \textbf{87}, 083507 (2013), arXiv: 1212.3611.


\bibitem{modesto3} C. Bambi \textit{et
al.}, Eur. Phys. J. C {\bf 74}: 2767 (2014), arXiv: 1306.1668.

\bibitem{modesto2} G. Calcagni \textit{et al.}, Eur. Phys. J. C {\bf 74}, 2999 (2014), arXiv:1306.5332.


\bibitem{efimovquantization} G. V. Efimov, Int. J. Theor. Phys. {\bf 10}, 19 (1974)
G. V. Efimov, Commun. Math. Phys. {\bf 7}, 138 (1968)


\bibitem{efimovunitarity} V. A. Alebastrov, G. V. Efimov, Commun. Math. Phys.
{\bf 31}, 1-24 (1973).

\bibitem{efimovcausality} V. A. Alebastrov, G. V. Efimov, Commun. Math. Phys. {\bf 38},
 11 (1974)

\bibitem{efimovqed} G. V. Efimov, M. A. Ivanov, O. A. Mogilevsky,
Ann. Phys. {\bf 103}, 169-l E4 (1977); G. V. Efimov and K.
Namsrai, Teor. Mat. Fiz. {\bf 22}, 186 (1975); G. V. Efimov, Ann. Phys. {\bf 71}, 466
(1972)

\bibitem{tomb} E. T. Tomboulis, hep-th/9702146.

\bibitem{biswasscalar} S. Talaganis, T. Biswas, A. Mazumdar, Classical 523
Quantum Gravity 32, 215017 (2015), arXiv:1412.3467[hep-th].

\bibitem{Staro}
A. A. Starobinsky, Phys. Lett. B {\bf 91}, 99 (1980); S. V. Ketov
and A. A. Starobinsky, JCAP {\bf 1208}, 022 (2012), arXiv: 1203.0805.

\bibitem{planck} Planck 2015 results, XIII, Cosmological parameters, Planck Collaboration: P. A. R. Ade et al.,
arXiv:1502.01589; Planck 2015, XX, Constraints on
inflation, Planck Collaboration: P. A. R. Ade et al.,
arXiv:1502.02114.

\bibitem{fieldred} N. Barnaby, Nucl. Phys. B {\bf 845}, 1 (2011), arXiv: 1005.2945;
A. Pais, G. E. Uhlenbeck, Phys. Rev. {\bf 79}, 145 (1950).

\bibitem{Itzkynson} C. Itzykson, J. B. Zuber, \textit{ Quantum Field Theory}, McGraw-Hill
Inc. (1980).

\bibitem{Spal} A. Smailagic, E. Spallucci, J. Phys. A36, L467 (2003), hep-th/0308193.

\bibitem{Lyth} D. Lyth, A. Liddle, \textit{The primordial density
perturbation}, Cambridge University Press (2009).

\end{thebibliography}
\end{document}